ARTICLE TYPE

# Momentum transfer coefficient constraints for the 2024 PDC25 Hypothetical Asteroid Impact Scenario

Christoph M. Schäfer*[1] | Uri Malamud*[2] | Tamir Manoach[2] | Hagai B. Perets[2]

[1]Department of Computational Physics, Institute for Astronomy & Astrophysics, University of Tübingen Germany

[2]Department of Physics, Technion - Israel Institute of Technology Israel

**Correspondence**
Christoph Schäfer
christoph.schaefer@uni-tuebingen.de
Uri Malamud urimala@physics.technion.ac.il
*These authors contributed equally to this work.
Schäfer - code development ; Malamud - initiative

**Abstract**

In Epoch 2 of the 2024 PDC25 Hypothetical Asteroid Impact Scenario, an asteroid is confirmed to be on a collision course with the Earth, and its size and surface composition have been well characterized via a flyby mission. A kinetic impactor deflection strategy is the most technologically mature path in order to mitigate this threat. Our goal is to constrain the possible range in momentum transfer coefficients, with implications for the number of impactors and the disruption risk. We conduct a series of numerical simulations, using a shock physics smoothed particle hydrodynamics code, in which we vary the impact velocity, cohesive properties and physical properties (mass / porosity) of the target asteroid. Given a judiciously chosen impactor mass, we show that the momentum transfer coefficient range is capable of a moderate-to-large enhancement of the asteroid deflection, yet keeps the disruption risk firmly at bay. These results are generally unique in having higher impact velocities compared to most previous studies.

**KEYWORDS**

planetary defense – smoothed particle hydrodynamics – Solar System asteroids

## 1 | INTRODUCTION

When detecting a potentially hazardous asteroid, a decision must be taken as to the best strategy for mitigating this threat. One must account for various parameters, including the asteroid's trajectory, its size, composition and structural properties.

In order to be ready when a real threat is presented, agencies around the world work jointly in preparation [16]. Exercises constitute as an important part of this exploratory process [6], and the latest such exercise is the 2024 PDC25 Hypothetical Asteroid Impact Scenario. In epoch 1 of this detailed exercise [†], a new near-Earth hypothetical asteroid (a.k.a. 2024 PDC25) is discovered on June 5, 2024 by the Catalina Sky Survey. Follow-up observations determine it has some probability of impacting the Earth. Initially, very little is known about its physical properties. In epoch 2 of this exercise[‡], massive subsequent astrometric observations are collected. By September 29, 2025, it is confirmed that the asteroid is on an Earth impact trajectory. Additionally, following the recommendation of the Space Mission Planning Advisory Group (SMPAG) after epoch 1 of the exercise, a fast flyby reconnaissance mission is launched in 2027, encountering the asteroid on April 12, 2028. The reconnaissance is able to greatly reduce physical uncertainties. The asteroid's effective diameter (150 ± 5 m) becomes well constrained, and its composition is determined to be S-type. However, its mass remains largely uncertain (2-7 $10^9$ kg).

The general strategies for dealing with any asteroid threat [9] are slow-push methods (gravity tractor, ablation by solar mirrors or lasers, ion beam, etc.) or impulsive methods such as nuclear explosive detonations (NED) or kinetic impacts (KI). In the context of 2024 PDC25, we focus on the kinetic impact strategy, which is the most mature and technologically tested strategy, following the success of the DART mission [8, 35]. It is also highly adequate for relatively small asteroids with sufficient warning time prior to impact. While several strategies may also be used in concert, here we will only address aspects related to kinetic impacts.

---

[†] https://cneos.jpl.nasa.gov/pd/cs/pdc25/index.html

[‡] https://cneos.jpl.nasa.gov/pd/cs/pdc25/epoch2.html

Common to all impulsive methods, the outcomes might vary between pushing the asteroid off its course intact (deflection) to dispersing it into multiple fragments (disruption). Deflection is usually favored over disruption, since in the latter case, the original asteroid is fragmented into a cloud of multiple smaller pieces, potentially extending the problem. If one were to consider disruption, one would favor dispersing the asteroid into the smallest pieces possible, over the largest volume possible. However, the uncertainties involved, make disruption a risky outcome, and typically one might want to avert that risk entirely, if possible, designing a mission that confidently avoids it. Here we adopt a similar approach and aim for a decisive deflection outcome.

Following the success of the DART mission, we now have a better assessment of the cohesive properties of typical rubble pile asteroids [26]. We know that kinetic impactors work, but we still need to establish that they work safely for other mission designs. In that regard, 2024 PDC25 makes an interesting comparison. It has a similar size to DART's target – Dimorphos – but its still *highly* uncertain mass even after the flyby mission of epoch 2, implies a range of possible porosities and cohesion properties. Additionally, typical impact velocities in potential 2024 PDC25 KI mission designs are up to about twice that of DART's velocity (of 6.1 km s$^{-1}$).

Given the focus placed on the DART mission in the last decade, indeed a large portion of the momentum transfer literature considers DART-like impactor speeds (E.g., [31, 23, 21, 22, 24, 33, 18, 25, 19]). Some studies did also consider higher velocities [37, 13, 3], however, while some simulations exceeded an impact velocity of 10 km s$^{-1}$, the cohesion they assumed was typically much higher than in the aforementioned investigations. Therefore, the combination of higher impactor velocities than 10 km s$^{-1}$, and the wide range in cohesive and porous properties of the target, make this study a unique and significant contribution. Under this range of conditions, calculating the lower and upper limits for the coefficients of momentum transfer, which are determined by the ratio of the total momentum (including the recoil of the ejecta) to that of the impacting spacecraft alone, is the paper's main goal. It will help place practical specific limits for the 2024 PDC25 exercise, because it could constrain how much deflection can be achieved, and if disruption can be avoided. At the same time, it would also broaden and strengthen the existing body of literature in this largely unexplored region of the impact parameter space.

## 2 | METHODS

In this section, we present the material and methods of our study. We briefly revisit topics that were explained in past works, and elaborate on their significance in our study, such as the Tillotson equation of state (EOS), the material strength model, the crush-curve for the $p-\alpha$ porosity model and the momentum transfer coefficient analysis. We used the open source smoothed particle hydrodynamics (SPH) code `miluphcuda` [27, 28, 29] for all conducted simulations. The `miluphcuda` code was successfully validated against experimentally derived results of momentum enhancement factors and benchmarked against other hydrocodes [18].

### 2.1 | Equations of state

For solid consolidated material, Tillotson is a widely used non-linear EOS for simulating high-velocity impacts and was first introduced by [36]. It can be applied over a wide range of physical conditions, while being computationally simple. It provides rudimentary distinction between the solid and vapour phase but lacks a consistent treatment of phase changes. In the case of compressed regions ($\varrho \geq \varrho_0$) and an energy $u$ that is lower than the energy of incipient vaporization $u_{iv}$, the EOS for the material pressure $P$ reads

$$P = \left(a_T + \frac{b_T}{1 + u/(u_0\eta^2)}\right)\varrho u + A_T\xi + B_T\xi^2, \tag{1}$$

with $\eta = \varrho/\varrho_0$ and $\xi = \eta - 1$. In case of expanded material ($u$ greater than the energy of complete vaporization $u_{cv}$) it reads

$$P = a_T\varrho u + \left(\frac{b_T\varrho u}{1 + u/(u_0\eta^2)} + A_T\xi\exp(-\beta_T(\eta^{-1} - 1))\right) \\ \times \exp(-\alpha_T(\eta^{-1} - 1)^2), \tag{2}$$

where $\varrho_0$, $A_T$, $B_T$, $u_0$, $a_T$, $b_T$, $\alpha_T$, and $\beta_T$ are material dependent parameters. In the partial vaporization regime ($u_{iv} < u < u_{cv}$), the pressure is linearly interpolated between the two values obtained via eq.(1) and eq.(2), respectively. We use the Tillotson EOS for modeling the asteroid solid matrix material, and the simpler, isothermal Murnaghan EOS for the kinetic impactor. The

latter reads

$$P = \frac{K}{n}\left(\left(\frac{\varrho}{\varrho_0}\right)^n - 1\right), \tag{3}$$

where $K$ denotes the bulk modulus, $\varrho_0$ the reference density in equilibrium and $n$ is a parameter. See Tables 1 and 2 for the values of the target and impactor, respectively.

## 2.2 | Porosity $P-\alpha$ model

Small asteroids in the Solar system are porous objects, as inferred from bulk density and composition measurements (e.g., [1]), which requires an additional modeling approach. Thus, for the porosity treatment, we implement the so called $P-\alpha$ model [10, 5], for which the pores are much smaller than the spatial resolution and cannot be modelled explicitly. Here, the total change in the volume depends on both the compaction of the pore space and the compression of the solid material which constitutes the matrix. The dependence is expressed in terms of the porous material pressure $P$ as $P_{solid}/\alpha$, where $\alpha$ is called the distention parameter, and is the ratio between the solid matrix material and the porous material densities, and relates to the porosity $\psi$ via $\psi = 1 - 1/\alpha$. As mentioned in the previous Section 2.1, for the solid matrix material we use the Tillotson EOS in order to obtain $P_{solid}$. Since in typical applications of planetary defense impacts the peak pressure is such that solid material compression is negligible, the key relation in the model becomes the dependence of $\alpha$ on the pressure, which is also known as the crush curve.

## 2.3 | Crush curve for porosity

The classic implementation by [12] uses a parametric quadratic mathematical function for the distention $\alpha(P)$. In [19], this choice was compared with both past and new empirical measurements, and it was suggested that for a wide range of potential pressures and/or initial porosities of the target asteroid, a more universal, non-parametric power law crush curve, might be a more suitable choice.

Here the pressure and the volume filling factor $\phi$ are linked by the following relation:

$$\phi = a'P^{b'}, \tag{4}$$

with the two constants $a'$ and $b'$. The relation between the distention and the volume filling factor is given by $\alpha = 1/\phi$. The elastic regime for the crush curve is determined from $a'$ and $b'$ and the initial distention $\alpha_0$ of the material via

$$P_e = \left(\frac{1}{a'\alpha_0}\right)^{1/b'}. \tag{5}$$

The implementation ensures that all deformations with a pressure below the $P_e$ limit do not lead to any compaction.

## 2.4 | Plasticity model

The plasticity model in our simulations follows the description in [18], using the Lundborg yield strength

$$Y(P) = Y_0 + \frac{fP}{1 + \frac{fP}{Y_M - Y_0}}, \tag{6}$$

where $Y_M$, $Y_0$ and $f$ are material dependent variables and denote the yield strength at infinite pressure, the cohesion, and the tangent of the angle of internal friction, respectively. For negative pressure, we decrease the strength with slope 1, i.e. the yield strength is zero for $P = -Y_0$. The values for the required parameters in the impactor and target are given in Tables 1, 2 and 3.

**TABLE 1** Material parameters for the impactor.

| Elastic properties | parameters from [18, 19] |
|---|---:|
| Bulk modulus $K$ [GPa] | 7.6 |
| Shear modulus $\mu$ [GPa] | 0 |
| Angle of internal friction $\alpha$ [deg] | 0.573 |
| Yield stress $Y_M$ [GPa] | 1 |
| Cohesion $Y_0$ [GPa] | 0.13 |
| **Murnaghan EOS** | |
| Reference density $\varrho_0$ [kg/m³] | 1000 |
| $n$ | 1 |

**TABLE 2** Material parameters for the target.

| Elastic properties | parameters from [18, 19] |
|---|---:|
| Bulk modulus $K$ [GPa] | 128 |
| Shear modulus $\mu$ [GPa] | 81.6 |
| Angle of internal friction $\alpha$ [deg] | 37.586 |
| Yield stress $Y_M$ [GPa] | 1 |
| Cohesion $Y_0$ | varied, see Table 3 |
| **Porosity parameters** | for pressure in MPa [19] |
| a' [(MPa)$^{-b'}$] | 0.41 |
| b' | 0.09 |
| **Tillotson EOS** | parameters for olivine [20] |
| Reference density $\varrho_0$ [kg/m³] | 3500 |
| $A_T$ [GPa] | 131 |
| $B_T$ [GPa] | 49 |
| $a_T$ | 0.5 |
| $b_T$ | 1.4 |
| $\alpha_T$ | 5.0 |
| $\beta_T$ | 5.0 |
| $u_0$ [J kg$^{-1}$] | $5.5 \times 10^8$ |
| $u_{iv}$ [J kg$^{-1}$] | $4.5 \times 10^6$ |
| $u_{cv}$ [J kg$^{-1}$] | $1.4 \times 10^7$ |

## 2.5 | Coefficient of momentum transfer analysis

In order to quantify the effect of the kinetic impact, we analyze how much momentum is transferred from the impactor to the target. The efficiency of momentum transfer from the impactor on the target is known as the $\beta$-factor (hereafter $\beta$ for short). We use the standard, simplified definition for the impact direction antiparallel to the surface normal at the point of impact and neglect any asymmetry in the ejecta plume. It can be expressed in terms of the impactor momentum $p_{\text{imp}}$, ejecta momentum $p_{\text{ej}}$ and the total change of momentum of the asteroid $\Delta p_{\text{ast}}$ (E.g., [9]) as

$$\beta = \frac{p_{\text{imp}} + p_{\text{ej}}}{p_{\text{imp}}} = \frac{\Delta p_{\text{ast}}}{p_{\text{imp}}} \approx \frac{m_{\text{ast}} \Delta v_{\text{ast}}}{m_{\text{imp}} v_{\text{imp}}}; \tag{7}$$

$$\Leftrightarrow \quad \Delta v_{\text{ast}} \approx \frac{m_{\text{imp}} v_{\text{imp}}}{m_{\text{ast}}} \beta, \tag{8}$$

where $m_{\text{ast}}$ and $m_{\text{imp}}$ denote the respective masses of the asteroid and the impactor, $v_{\text{imp}}$ is the relative impact velocity and $\Delta v_{\text{ast}}$ denotes the change in the orbital velocity of the asteroid. A factor of $\beta \approx 1$ indicates that the ejecta plume has no major contribution on the deflection efficiency, while $\beta > 2$ means that the momentum contribution from the crater ejecta is larger than from the impactor. For the numeric calculation of $\beta$ we use the same procedures that were used in the [18] study based on the predecessor work [13]. All the particles with ejection velocities exceeding the asteroid's escape velocity $v_{\text{esc}}$ are summed

for the ejecta momentum

$$p_{\text{ej}} = \sum_i m_i \, v_i, \tag{9}$$

where $m_i$ denotes the mass of ejected particle $i$ and $v_i$ its velocity in the vertical direction. To determine the $\beta$-factor, we follow this procedure as a postprocessing step and compute $\beta$ for several snapshots in the last part of the simulation time to make sure to obtain the converged value.

The escape velocity is given by

$$v_{\text{esc}} = \sqrt{\frac{2G m_{\text{ast}}}{R_{\text{ast}}}} \tag{10}$$

where $R_{\text{ast}}$ denotes the asteroid's effective radius.

## 2.6 | Model realizations

The variations in the properties of the hypothetical asteroid are given in epoch 2 of the 2024 PDC25 exercise. The effective radius of the asteroid is very well constrained (∼ 75 ± 2 m). Since the possible asteroid masses (and in turn densities) are not well constrained, a statistical distribution is provided instead. [§]. For convenience, we limit ourselves to model the low (5th percentile), intermediate (50th percentile), and high (95th percentile) masses in this statistical distribution (Table 3). In Section 4 we discuss why modeling more cases (like the 25th and 75th percentiles) would not alter our final conclusions. We aim to obtain sufficiently different porosities in our investigation, since porosity is one of the two main parameters that affects the momentum transfer, as more porous objects experience greater compression and in turn reduce the ejecta (and thus $\beta$). The other main parameter affecting $\beta$ is the assumed cohesion. Unlike porosity, which is obtained directly from knowledge of the material and bulk porosity, the cohesion has to be assumed. Our principal goal in the paper is to place lower and upper limits, which then translate to lower and upper limits for $\beta$.

We further choose our solid matrix target material. Hence, the porosity $\psi$ results directly from the bulk density $\rho$ and the material specific density $\varrho$ of the target as $\psi = 1 - \rho/\varrho$ (see the porosity row in Table 3). Here we choose the material to be olivine, which is both similar to the S-type surface composition identified in the epoch 2 flyby mission, and also has a specific density of 3500 kg/m$^3$, which approximately matches the maximal density of the asteroid in the statistical distribution. Since olivine's specific density is slightly higher than the 95th percentile's density, our simulations of the highest mass asteroid are only slightly porous. Finally, olivine has well known and often used Tillotson EOS parameters (Table 2).

For our model cohesion values we consider a total of 7 cases. We have lower and upper limit cohesion values for each of the 3 asteroid masses, and an additional intermediate cohesion for the high mass asteroid, since in the latter case the range in potential cohesion values is greater. Our considerations are as follows. The low and intermediate mass asteroids have bulk porosities that are probably compatible with rubble pile asteroids ($\psi$>36.7%). In recent years, we have learned from planetary space missions to rubble pile asteroids that their cohesion is much lower than was assumed in previous decades, and might be only around several Pa or even sub Pa (e.g., [15, 25]). Somewhat less direct estimations consider bulk cohesion values up to hundreds of Pa, e.g. as reviewed in the introduction section of [11]. Some newer studies also postulate that even for the DART impact, which was previously interpreted as consistent with Pa scale cohesions, values up to hundreds of Pa might also be plausible, especially given uncertainties in various model relations such as the internal friction or the crush curve [30, 34]. Thus, for our low and intermediate mass asteroids (for whom bulk porosity is compatible with a rubble pile structure) we choose similar values. For the low mass asteroid, the bulk porosity is nearly 60%, suggesting a combination of large macro and micro porosity. In these circumstances we estimate the lower and upper values for the cohesion to be 0.1 and 100 Pa, respectively. For the more typical bulk porosity of the intermediate mass asteroid (around 36%) we select 1 and 1000 Pa. The high mass asteroid, however, has a very low porosity around 10% only. Such an object is nearly consolidated, and one might expect a consolidated body to have a much larger cohesion. As an upper limit, we select 1 MPa as the cohesion, being closer to the cohesion of meteorites. However, the small porosity might also represent fracture in the asteroid, that tends to lower bulk cohesion considerably. In addition, intact material can become damaged in the impact. We did not include fracture in our simulations explicitly, since this would have applied only to the high mass / low porosity model realization, and we preferred to use the same simulation setup in all cases. Nevertheless, the lower and intermediate cohesion values we set may also be viewed as indicative of fully damaged material. We select for the latter 1 and 1000 Pa, respectively.

---

[§] https://cneos.jpl.nasa.gov/pd/cs/pdc25/2024pdc25_epoch2_physical_properties.zip

In order to complete the above scheme, we select two impact velocities which approximately represent the lower (10 km s$^{-1}$) and upper (12.5 km s$^{-1}$) limit velocities achieved during the impact, given the orbit of 2024 PDC25. The impact angle in all simulations is assumed to be normal to the surface. The mass of the impactor is fixed at 200 kg, since a priori estimations (and the results now obtained in this study) showed that this mass will be sufficiently low to avoid disruption of 2024 PDC25.

In total, we have performed 2x7=14 simulations.

## 2.7 | Code setup

In this study We used the open source smoothed particle hydrodynamics (SPH) code `miluphcuda` [27, 28, 29]. We performed simulations based on the parameters and setup from the DART impact benchmark study of [18], with a modified crush curve based on [19], specifically suited to treat a wide range of porosities, as discussed in Section 2.3.

This setup consists of SPH particles distributed in a spherical impactor and a flat, half-sphere target. To increase the accuracy at the impact point, the number of target particles decreases radially from the impact point. The target particle number density in the impact point matches the number density of the impactor to avoid numerical issues.

The resolution in all simulations was about 500 000 SPH particles. The numerical parameters were set to match the parameters in [18], where resolution convergence tests showed the selected resolution to be sufficient, and therefore we do not repeat convergence tests here. The simulations ran on the DGX cluster at the Technion, with each simulation running on a single Nvidia A100 GPU. The runtime per simulation was between 1 and 2 weeks.

We continued each SPH simulations until we achieved temporal convergence for $\beta$. Characteristically, convergence takes about 10 s, however the convergence time is not identical for every simulation, the full range in our entire simulation suite (discussed next) being 5-22 s. Note that our goal for this paper was only to calculate $\beta$, and therefore we did not simulate the crater formation to completion, which is far more demanding and exceeds the scope of this study.

## 3 | RESULTS

The results of the various simulation realizations described in Section 2.6 are shown in Table 3, along with the simulation parameters. All other parameters being equal, there are several general trends in the simulation data. The momentum transfer coefficient $\beta$ anti-correlates with both the cohesion and the porosity. On the other hand, $\beta$ correlates with the impact velocity, as would be expected. However, the ratio $\beta_{12.5}/\beta_{10}$ is not constant. It is seen to grow from left to right, i.e. with increase in both the cohesion and the density. At the limit of a nearly consolidated, largely cohesive target, this ratio increases with nearly linear proportion to the impact velocity.

Interestingly, $\beta$ is around three times higher in the case of the impact in the 11.8% porous asteroid with cohesion 1 Pa compared to the impact in the 57.57% porous asteroid, even when the cohesion is ten times lower 0.1 Pa in the latter case. This is understood from the greater energy that goes to compression in more porous targets, which in turn implies less energy to the ejecta. Figure 1 demonstrates effectively the dependency of $\beta$ on the ejecta production efficiency - the ejecta curtain in the high $\beta$ case includes much more material compared to the low $\beta$ case.

Using Equation 8, we can calculate $\Delta v$. Using Equation 10, we calculate the escape velocity. The ratio between them is, however, non trivial, as will be discussed next.

## 4 | DISCUSSION

### 4.1 | Deflection versus disruption

An initial planning step in setting up the simulations was to select the impactor mass in such a way that would confidently avoid the asteroid's disruption. While $\beta$ depends primarily on the properties of the asteroid, such as its cohesion and porosity, Equation 8 shows that $\Delta v$ is proportional to the impactor's mass. By taking a relatively small mass, and given a strong upper limit for $\beta$, we can guarantee keeping the ratio between $\Delta v$ and $v_{esc}$ sufficiently low, having knowledge of the asteroid's mass and impact velocity. As already known from prior studies, a low ratio leads to a deflection, and avoids the risk of a disruption. One of the main purposes of the current study was to find out the upper limit for $\beta$, rather than simply make an educated (yet speculative) guess, as we have done prior to setting up the simulations.

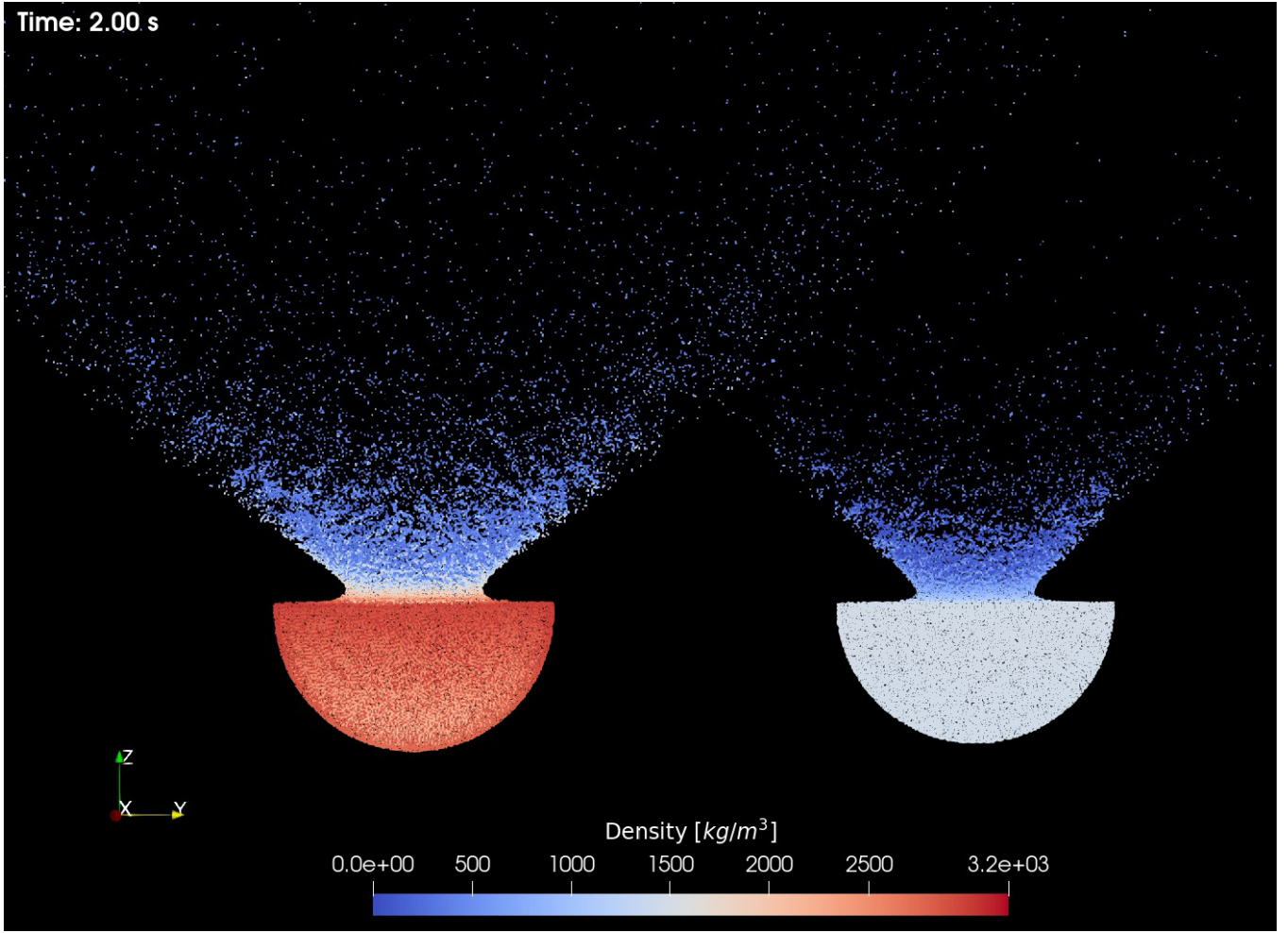

**FIGURE 1** Ejecta curtains for the highest ($\beta$ = 7.92, left) and lowest ($\beta$ = 1.93, right) measured $\beta$ factor for impact velocity 12.5 km s$^{-1}$ after a simulation time of 2 s. Colour coding indicates the densities of the SPH particles. Please note the two different densities of the asteroids due to the different porosities of 11.8% (left) and 57.57% (right), respectively.

The results presented in Section 3 show that $\beta$ increases with the asteroid's mass (porosity decreases as the radius remains nearly constant). However, Equation 8 shows that, mathematically, while the numerator increases $\Delta v$ in that case, the asteroid's mass in the denominator has the opposite effect of decreasing it. In combination, these two effects tend to cancel out, and $\Delta v$ among the various mass percentile are much more similar than the $\beta$ values. An even smaller difference exists in $\Delta v/v_{esc}$, as $v_{esc}$ goes like the square root of the asteroid's mass. The two highest values of $\Delta v/v_{esc}$ (bold text in Table 3), occur for the most porous and least porous models, respectively, for the lower limit cohesion. In both cases the ratio is about 3.5% (3.9% and 3.3%, respectively).

The common benchmark for disruption risk is usually around $\Delta v/v_{esc}$ = 10% [4, 7]. This latter statement usually assumes that the $\Delta v/v_{esc}$ benchmark is independent on the asteroid properties. However, the disruption benchmark does depend on the asteroid properties, as shown by numerous studies (e.g., [17, 14, 25]). The problem is often formulated in terms of the energy of catastrophic disruption $Q_D*$ and not in terms of $\Delta v/v_{esc}$, although these descriptions are related. The specific impact energy [2] is

$$Q = \frac{v_{imp}^2}{2} \frac{m_{imp}}{m_{ast}}. \tag{11}$$

**TABLE 3** Parameters and results from all 14 simulations. The maximum value of the velocity difference in units of the escape velocity is approximately $\Delta v/v_{esc}$ = 3.5% (boldface in the table). All cases are well below the 10% disruption risk limit.

| Parameters | | | | | | | |
|---|---|---|---|---|---|---|---|
| Percentile | [%] | 5th | | 50th | | 95th | |
| Density $\varrho$ | [kg/m³] | 1485 | | 2229 | | 3087 | |
| Porosity $\psi$ | [%] | 57.57 % | | 36.3 % | | 11.8 % | |
| Radius $R$ | [m] | 73 | | 75 | | 77 | |
| Mass $m$ | [$10^9$ kg] | 2.42 | | 3.94 | | 5.90 | |
| Escape vel. $v_{esc}$ | [cm/s] | 6.652 | | 8.373 | | 10.116 | |
| Cohesion $Y_0$ | [Pa] | 0.1 | $10^2$ | 1 | $10^3$ | 1 | $10^3$ | $10^6$ |
| Results | | | | | | | |
| Impact velocity $v_{imp}$ = 10 km s⁻¹ | | | | | | | |
| Mom. transf. coeff. $\beta$ | | 2.4668 | 1.8363 | 3.1731 | 1.9168 | 6.7644 | 4.135 | 2.1991 |
| Velocity diff. $\Delta v$ | [cm/s] | 0.2038 | 0.1517 | 0.1611 | 0.0973 | 0.2291 | 0.1401 | 0.0745 |
| $\Delta v/v_{esc}$ | | 0.031 | 0.023 | 0.019 | 0.012 | 0.029 | 0.014 | 0.007 |
| Impact velocity $v_{imp}$ = 12.5 km s⁻¹ | | | | | | | |
| Mom. transf. coeff. $\beta$ | | 2.4937 | 1.9317 | 3.5538 | 2.2428 | 7.9197 | 5.1095 | 2.7167 |
| Velocity diff. $\Delta v$ | [cm/s] | **0.2576** | 0.1996 | 0.2255 | 0.1423 | **0.3354** | 0.2163 | 0.1150 |
| $\Delta v/v_{esc}$ | | 0.039 | 0.03 | 0.027 | 0.017 | 0.033 | 0.021 | 0.011 |

Thus, as our current impact velocity is up to approximately twice that of DART, while the DART impactor mass is about thrice that of our own simulated impactor, these effects cancel out. The specific impact energy in both cases is quite similar. In turn, building on the work of [22], the caution zone that underlines the gap between substantial disruption risk and mere reshaping of the asteroid, might reach a number of the order of a few. We conclude that if our maximal $\Delta v/v_{esc}$ stays below 3.5%, this marks a significant gap from the 10% benchmark value, and we can be more confident about a deflection outcome. Only in one case in the entire simulation suite do we obtain a value of 3.9%, which is sufficiently close to our 3.5% caution benchmark.

Table 3 also shows why the 3 simulated cases (percentiles 5, 50 and 95) were sufficient for demonstrating robust deflection. The maximal $\Delta v/v_{esc}$ values were obtained for the end cases, and thus more intermediate cases (e.g., percentiles 25 and 75) would not have altered our conclusions.

## 4.2 | Comparison with previous studies

The dependency of $\beta$ on porosity and cohesion was already found in several different studies (see, e.g., [13, 33]), and is directly linked to the higher production of ejecta for less porous and less cohesive asteroid material compared to highly porous objects with high cohesion.

The increase in $\beta$ with velocity is also demonstrated in past studies. Our result that this increase is much greater as one transitions to more consolidated and more cohesive targets, seems to have also been observed in past numerical and empirical studies [13, 3], including the more linear proportions in consolidated targets [38], although in the latter study the experimental velocities were lower than in our current numerical study.

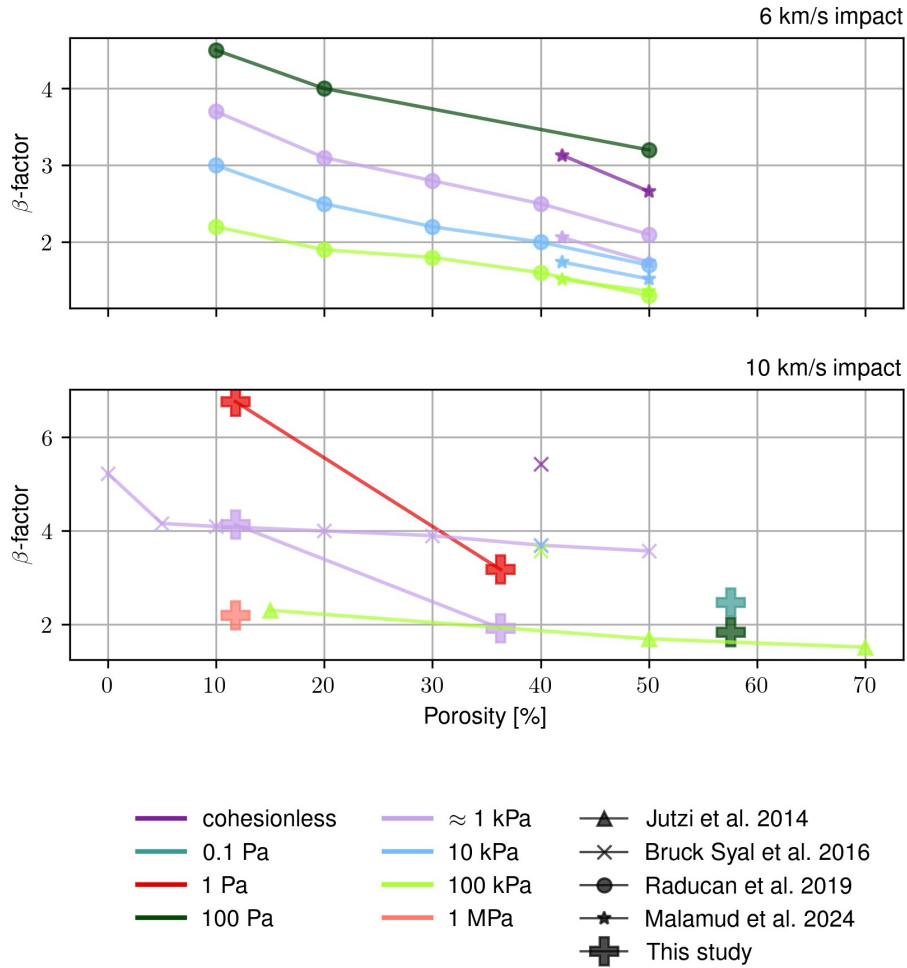

**FIGURE 2** Momentum enhancement factor $\beta$ for a DART-like impact velocity (top panel) and higher impact velocity of $10 \text{ km s}^{-1}$ (lower panel) from various studies [13, 3, 23, 19].

A direct comparison between studies is often challenging, since in reality not only that different codes are used, but also the details in the model constituent relations are not the same, thus affecting the simulation outcomes. With this caveat in mind, we would still like to broadly examine trends across various past studies, to the extent that we can (generally) compare similar parameters. Indeed, a couple of benchmarking studies that compared between different codes [32, 18], have shown a good agreement between impact results, and specifically for estimating $\beta$ (especially when care was taken to use consistent crushing behavior). Based on these past findings, we present our comparison in Figure 2.

Since some past studies already investigated the significance of porosity for the amount of ejecta, and in turn the $\beta$ coefficient, we likewise display the latter as a function of the former, and can compare our results. We note that the problem is multi-faceted, as $\beta$ depends not merely on porosity, but also on other parameters, and in particular the cohesion. Various studies used vastly different values for the cohesion, and therefore we add another layer of information, by plotting different cohesion values using a color scheme (see legend). Some previous studies happen to have chosen the same cohesion values for their simulations. For visual clarity, we therefore assign different symbols for data points that are attributed to different studies (see legend). In order to complete this scheme, we differentiate between two pivotal impact velocities. The top panel focuses on simulation results in previous studies which simulated DART-like impact velocities of $6 \text{ km s}^{-1}$, whereas the bottom panel focuses on simulation results (including this study) at a higher impact velocity of $10 \text{ km s}^{-1}$, in which two other previous studies have also performed

10 | Christoph M. Schäfer, Uri Malamud, Tamir Manoach, Hagai B. Perets

simulations. The reason we do not show a third panel for the upper limit impact velocity in this study of 12.5 km s$^{-1}$, is that there was no other previous study performed at this exact velocity.

The upper panel in Figure 2 emphasizes that while results obtained within different simulations are generally similar, and follow the same trend, certain mismatches are unavoidable, due to important differences that are not manifested in the plot. For example, the [19] study used a different crush curve for modeling the porous target material, which amounted to more compression and thus less ejecta and smaller $\beta$ values. At a similar porosity of 50%, the $\beta$ obtained for the [19] collisionless simulation was smaller than the $\beta$ obtained by [23] despite using a larger cohesion value of 100 Pa. Such differences are also observed in the bottom panel that newly compares the results of this study with previous ones. For example, the slope with porosity obtained by [3] is much shallower than the one obtained in the current study, for an identical choice of cohesion (1 kPa lines). Thus, both studies only agree on $\beta$ for a low porosity, where compression becomes more negligible. On the other hand, [13] obtained similar $\beta$ values as a function of the porosity compared to the current study, however there are significant differences in the cohesion values adopted between the studies. Such distinctions are already known to affect impact simulations results (e.g., as was nicely summarized in figure 2 of [33]).

Based on the bottom panel, and especially the results from the current study, it appears that at high velocities exceeding 10 km s$^{-1}$, the differences in $\beta$ among different cohesion values diminish much more as the porosity increases. In other words, cohesion seems pivotal for more consolidated targets, whereas highly porous targets are less sensitive to it.

## 5 | CONCLUSIONS

This study's findings amount to more than just the specific scenario of 2024 PDC25. We present some of the first ever simulations to comparatively investigate a wide range of porosities and cohesions, while using a new, parameter-independent crush curve, that allows for a more universal treatment of various target porosities. In addition, we perform simulations at impact velocities exceeding 10 km s$^{-1}$, which were rarely considered in past numerical simulations, due to the focus placed on slower DART-like impacts. Thus, this contribution enriches the current impact literature.

In the specific context of the PDC25 exercise, even the maximum velocity difference of 3.354 mm s$^{-1}$ for the least porous and most cohesive asteroids considered in our study, will not be sufficient to prevent the collision of 2024 PDC25 with Earth[¶] in a single deflection mission. Hence, a strategy of using more massive impactors and/or multiple kinetic impactors (which is beyond the scope of this study) has to be considered and planned.

Without prior knowledge about the potential momentum transfer coefficient in the impact, our simulations robustly suggest that for 2024 PDC25, using a more massive single impactor is not actually a possibility, as long as one wishes to confidently avoid the risk of disrupting the asteroid. At the same time, our simulations do indicate that a fast reaction on a potential asteroid threat as defined in the 2024 PDC25 scenario, using a low-mass kinetic impactor of 200 kg, will provide partial deflection as well as valuable insights into the material properties of the asteroid, while keeping $\Delta v / v_\text{esc}$ safely below the disruption risk limit. In turn, the knowledge gained in the aforementioned fast-reaction single kinetic impact, would be used in order to optimize impactor masses for subsequent kinetic impacts. While we would not be able to achieve full deflection with a single mission, mass optimization would be able to minimize the number of subsequent missions.

## 6 | ACKNOWLEDGEMENTS

We acknowledge the work by Christoph Burger who originally designed the setup that was used in predecessor studies. Part of this work was supported by the German Research Foundation (DFG) project number Ts 17/2–1. CMS acknowledges support by the High Performance and Cloud Computing Group at the Zentrum für Datenverarbeitung of the University of Tübingen, the state of Baden-Württemberg through bwHPC and the DFG through grant no INST 37/935-1 FUGG. UM and HBP acknowledge support from the Israeli Ministry of Science and Technology MOST-space grant and the Minerva Center for Life under extreme conditions. We acknowledge support from the Open Access Publication Fund of the University of Tübingen.

---

[¶] https://cneos.jpl.nasa.gov/nda/nda.html

https://www.sciencedirect.com/science/article/pii/S0734743X19308048.